\documentclass[conference,a4paper]{APSIPA2025}
\usepackage{amsmath}
\usepackage{amsfonts}
\usepackage{graphicx}
\usepackage{multirow}
\usepackage{threeparttable}
\usepackage{float}
\usepackage{hyperref}

\usepackage{geometry}
\usepackage{fancyhdr}

\begin{document}

\title{From Evaluation to Optimization: Neural Speech Assessment for Downstream Applications}

\author{
Yu Tsao\\    
        Research Center for Information Technology Innovation, Academia Sinica, Taipei, Taiwan \\
        E-mail: yu.tsao@citi.sinica.edu.tw

}

\maketitle

\begin{abstract}
The evaluation of synthetic and processed speech has long been a cornerstone of audio engineering and speech science. Although subjective listening tests remain the gold standard for assessing perceptual quality and intelligibility, their high cost, time requirements, and limited scalability present significant challenges in the rapid development cycles of modern speech technologies. Traditional objective metrics, while computationally efficient, often rely on a clean reference signal, making them intrusive approaches. This presents a major limitation, as clean signals are often unavailable in real-world applications. In recent years, numerous neural network–based speech assessment models have been developed to predict quality and intelligibility, achieving promising results. Beyond their role in evaluation, these models are increasingly integrated into downstream speech processing tasks. This review focuses on their role in two main areas: (1) serving as differentiable perceptual proxies that not only assess but also guide the optimization of speech enhancement and synthesis models; and (2) enabling the detection of salient speech characteristics to support more precise and efficient downstream processing. Finally, we discuss current limitations and outline future research directions to further advance the integration of speech assessment into speech processing pipelines.
\end{abstract}

\section{Introduction}
The development of advanced speech processing systems, ranging from enhancement and dereverberation to synthesis and conversion, has long been driven by the goal of improving speech quality and intelligibility for human listeners. Yet, progress toward this objective has been hindered by a persistent disconnect between the metrics used for algorithmic optimization and the complex, nuanced nature of human auditory perception. This perceptual gap stems from the limitations of traditional evaluation and optimization paradigms, which either depend on computationally convenient yet perceptually misaligned mathematical measures or on human-centered subjective tests that, while accurate, are costly, time-consuming, and difficult to scale. Recognizing this gap is critical to understanding the transformative role of modern neural speech assessment methodologies~\cite{cooper2024review}.

Neural speech assessment models have recently emerged as a prominent research focus and are increasingly integrated into a wide range of speech processing tasks. In speech enhancement, notable examples include 
DNSMOS~\cite{reddy2021dnsmos}, MOSA-Net~\cite{zezario2022deep}, PESQ-DNN~\cite{xu2023coded}, SpeechBERTScore~\cite{saeki2024speechbertscore}, VQScore~\cite{fu2024self}, Quality-Net~\cite{fu2018quality}, STOI-Net~\cite{zezario2020stoi}, and SpeechLMScore~\cite{maiti2023speechlmscore}.

For voice conversion and text-to-speech (TTS) systems, widely adopted speech assessment models, such as MOSNet~\cite{lo2019mosnet}, MBNet~\cite{leng2021mbnet}, NORESQA~\cite{manocha2021noresqa}, NORESQA-MOS~\cite{manocha2022speech}, 
LDNet~\cite{huang2022ldnet}, SSL-MOS~\cite{cooper2022generalization}, UTMOS~\cite{saeki2022utmos}, SOMOS~\cite{maniati2022somos}, LE-SSL-MOS~\cite{qi2023ssl}, and TTSDS2~\cite{minixhofer2025ttsds2}, have demonstrated effectiveness in numerous benchmark evaluations~\cite{huang2022voicemos, cooper2023voicemos, huang2024voicemos} and have been employed as evaluation metrics in various speech processing challenges~\cite{zhang2024urgent, blanco2023avse}.

More recently, multimodal neural speech assessment approaches have been proposed, incorporating additional information such as contextual cues~\cite{wang2025qualispeech} and visual signals~\cite{ahmed2025study} to improve accuracy and alignment with human perception. In parallel, emerging research has investigated the use of alternative biomarkers, such as physiological or cognitive indicators, as objective measures for predicting speech quality and listening effort~\cite{parmonangan2023common, hsin2025exploring}.

\begin{figure}
    \centering
\includegraphics[width=1.0\linewidth]{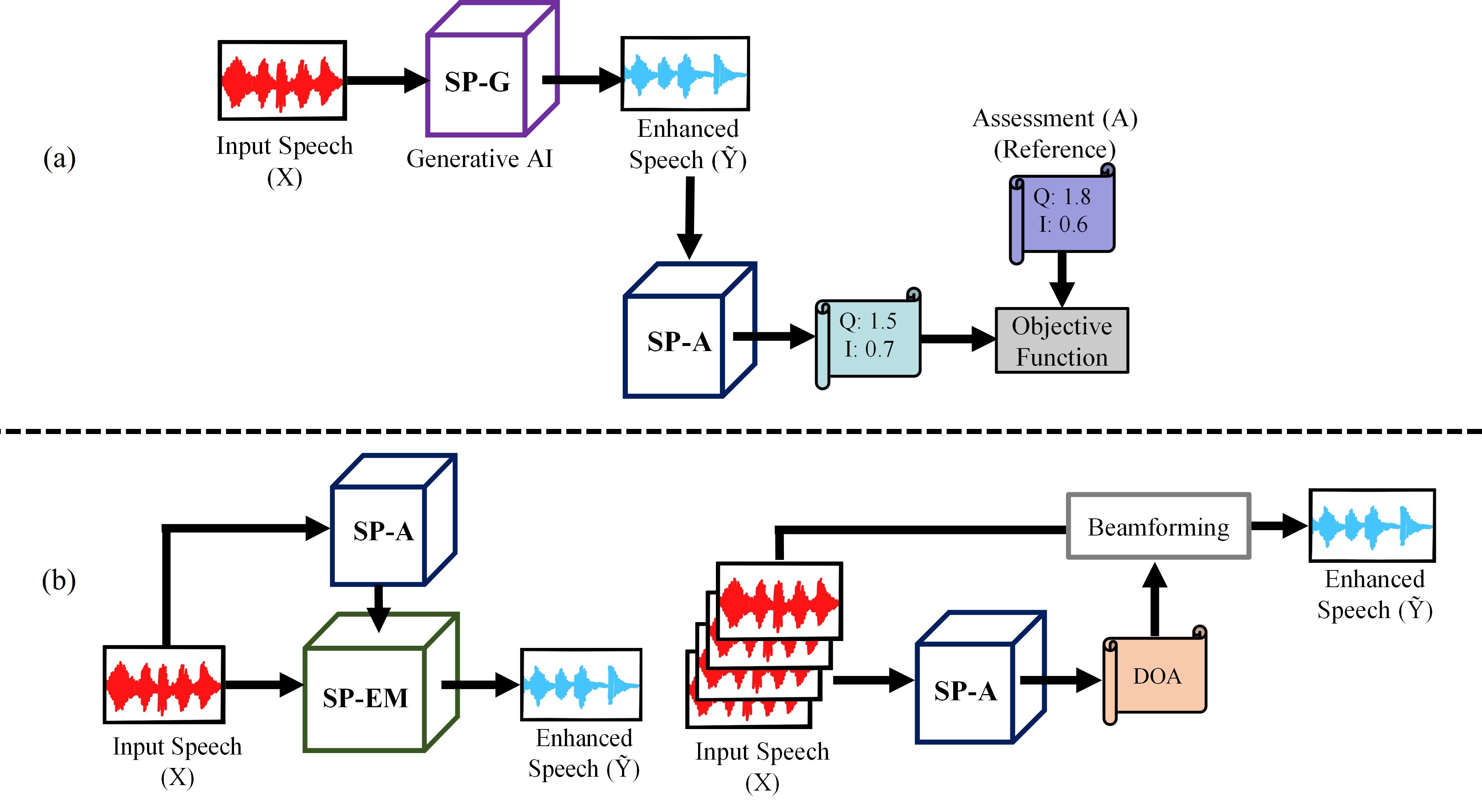}
    \caption{Neural speech assessment supports two major downstream applications: (1) serving as a differentiable perceptual proxy to guide the optimization of speech generation models, and (2) enabling the detection of key speech characteristics for more precise and efficient downstream processing. Left: model selection in an ensemble framework; Right: DOA estimation for beamforming. SP-G: speech generation model; SP-A: speech assessment model; SP-EM: speech generation using an ensemble system.}
    \vspace{-10pt}
    \label{fig:overall}
\end{figure}

Beyond their role as evaluation tools, neural speech assessment models have been shown to enhance the performance of speech processing pipelines~\cite{fu2019metricgan, nayem2021incorporating, xu2021deep, fu2022metricgan, xu2022deep, ting2023ians,  nayem2023attention, wang2024unsupervised, chao2024investigation, wu2025reexamining, lu2025explicit}. This review focuses on their application in downstream speech processing tasks, which can be broadly classified into two categories, as illustrated in Fig.~\ref{fig:overall}:

(1) Perceptually aligned and differentiable metrics – models that not only evaluate but also guide the training of speech enhancement and synthesis systems~\cite{fu2019metricgan, wang2024unsupervised, jiang2025mos}.

(2) Speech property identification – models that detect key speech characteristics, enabling more targeted and effective downstream processing~\cite{zhang2024urgent, ting2023ians, zezario2019specialized, zezario2021speech}.

For the first category, traditional signal-level loss functions, such as mean squared error (MSE) and mean absolute error (MAE), are computationally efficient and differentiable, making them suitable for training speech synthesis models. However, these measures correlate suboptimally with human perception and often produce over-smoothed, unnatural outputs. Subjective listening tests, such as MOS, provide the most accurate perceptual evaluation but are costly, time-consuming, and prone to biases, with limited discriminative power for top-performing systems.

Objective metrics like perceptual evaluation of speech quality (PESQ)~\cite{rix2001perceptual} and perceptual objective listening quality assessment (POLQA) ~\cite{gaoxiong2012perceptual}, better approximate perception than MSE and MAE, yet they struggle with non-differentiable, preventing their direct use as training objectives. In addition, these metrics are intrusive, meaning they require a reference signal for comparison when assessing the target speech.

These limitations have driven the development of neural speech assessment models: non-intrusive, differentiable, and data-driven perceptual proxies trained to approximate human judgments or established metrics. Importantly, these models can be integrated into training pipelines as perceptually aligned loss functions, enabling direct optimization toward human-perceived quality. Frameworks such as MetricGAN~\cite{fu2019metricgan} and its extensions~\cite{fu2022metricgan, fu2021metricgan+, cao2022cmgan, close2022metricgan+, cheng2023speech, shin2023metricgan, hou2023convolutional, mai2025metricgan+} exemplify this paradigm shift, turning evaluation from a passive, post-hoc process into an active, perception-driven component of model learning.

More recently, neural assessment–driven optimization has emerged not as a mere incremental enhancement to existing tasks, but as a foundational technology that enables entirely new capabilities and addresses long-standing challenges in speech processing. By extending the concept of a learned perceptual loss function, researchers have expanded the boundaries of what is possible—enabling fully unsupervised learning from real-world data, direct optimization for subjective human preferences, and intelligent, perception-aware control of complex audio systems.

For the second category, neural speech assessment models can effectively characterize the properties of speech signals, enabling tasks such as filtering out low-quality speech~\cite{zhang2024urgent} or selecting the most suitable speech processing model for a given input. In~\cite{zezario2019specialized, zezario2021speech}, a speech enhancement framework was proposed that employs an ensemble of specialized models, guided by Quality-Net—a pre-trained, non-intrusive neural network that predicts PESQ scores without requiring a clean reference. This ensemble-based strategy has demonstrated superior generalization performance compared to a single, general-purpose model.

Finally, neural speech assessment can be integrated with traditional acoustic beamforming systems to further enhance performance. Accurate direction-of-arrival (DOA) estimation is critical in beamforming but remains challenging, particularly under very low signal-to-interference ratio (SIR) conditions. In~\cite{ting2023ians}, STOI-Net, a non-intrusive neural speech assessment model that predicts short-time objective intelligibility (STOI)~\cite{taal2011algorithm} scores, is used to estimate the DOA by evaluating predicted scores for speech signals from a set of candidate angles. The angle yielding the highest STOI score is selected as the target DOA, after which a beamforming algorithm is applied to focus on that direction and extract the target speech.


\section{Neural Speech Assessment Models as Differentiable Perceptual Proxies}

Neural speech assessment models are learning to ‘listen’ and to ‘evaluate’ speech in alignment with human judgment, and, being fully differentiable, can be seamlessly integrated into model training pipelines.

\subsection{The Surrogate Concept: Differentiable Mirrors of Black-Box Metrics}

The core concept is to train a neural network to approximate a complex, non-differentiable function. Once trained, this network—serving as a “surrogate”—can replace the original function as a differentiable loss. A notable example is Quality-Net~\cite{fu2018quality}, a Bidirectional Long Short-Term Memory (BLSTM) model, designed as a differentiable proxy for the PESQ metric. Quality-Net takes the spectrograms of a degraded signal and its clean reference as input, and is trained to predict the PESQ score that the original algorithm would produce. By learning this mapping, it transforms the ‘black-box’ PESQ function into a ‘white-box’ that provides usable gradients for backpropagation, enabling a speech enhancement model to be fine-tuned to directly maximize its predicted PESQ score.

\subsection{Advancing the Paradigm: From Intrusive Metrics to Human Judgments}
\label{sec:Advacing}

While creating surrogates for intrusive metrics like PESQ was a significant breakthrough, the need for a clean reference signal remained a key limitation. The next stage of advancement focused on developing models capable of operating non-intrusively and, ultimately, on training models directly from human subjective ratings—bypassing traditional objective metrics altogether.

\begin{itemize}
    \item Non-Intrusive Proxies: Models such as Quality-Net~\cite{fu2018quality} and STOI-Net~\cite{zezario2020stoi} were designed to predict metric scores using only the degraded signal. Quality-Net estimates PESQ scores, while STOI-Net predicts STOI scores, each by implicitly learning the acoustic characteristics critical to speech quality and intelligibility, respectively. Such non-intrusive assessment is essential for real-world, real-time applications where a clean reference signal is unavailable.

    \item Learning Directly from Humans: The most advanced neural assessors are trained on large-scale datasets of human-rated speech. For example, DNSMOS~\cite{reddy2021dnsmos} is trained on extensive collections of noisy clips, each annotated by human listeners with MOS ratings for signal quality, background noise, and overall quality. Similarly, MaskQSS~\cite{wang2024unsupervised} is a specialized model designed to predict the MOS of speech distorted by face masks, trained on a custom database of human-rated, mask-recorded speech. This human-in-the-loop approach enables models to learn a direct mapping from acoustic features to human preference, capturing perceptual subtleties overlooked by traditional metrics and facilitating the development of highly specialized assessors for specific application domains.
\end{itemize}

\subsection{From Assessor to Optimizer: The MetricGAN Framework}

The true strength of differentiable neural assessment lies in its ability to be integrated into the training process as an active, adaptive loss function. The MetricGAN framework exemplifies this paradigm, repurposing a Generative Adversarial Network (GAN)~\cite{goodfellow2020generative} architecture to directly optimize for any black-box evaluation metric. 

The MetricGAN framework consists of two parts:
\begin{itemize}
\item Generator (G): The speech enhancement model being optimized.
\item Discriminator (D): A neural speech assessment model that serves as a surrogate for the target metric (e.g., PESQ). Rather than classifying real vs. fake, it is trained to predict the score that the target metric would assign to a given speech sample.
\end{itemize}

The training follows a minimax game in which the discriminator learns to become an increasingly accurate predictor of the target metric for the specific types of speech produced by the generator. In turn, the generator receives gradients from the discriminator, guiding it to produce outputs that maximize the predicted score. The discriminator thus serves as a learned, adaptive loss function, continuously refining its understanding of the quality landscape based on the generator’s evolving artifacts. This provides a more robust and contextually relevant training signal than a static, pre-trained model.

The MetricGAN+~\cite{fu2021metricgan+} framework introduced several engineering enhancements to stabilize training and improve performance. These include training the discriminator on the original noisy speech (in addition to clean and enhanced speech) to provide stronger reference anchors, employing an experience replay buffer to prevent catastrophic forgetting, and integrating a learnable, per-frequency sigmoid activation in the generator for more flexible noise suppression. Together, these improvements yielded significant performance gains, underscoring the strong relationship between the quality of the learned loss function and the resulting perceptual quality of the output.

\subsection{The Unsupervised Revolution with MetricGAN-U}

A major bottleneck in supervised speech enhancement is the reliance on large, parallel corpora of noisy and clean speech, which are costly to produce. The MetricGAN-U (Unsupervised) framework~\cite{fu2022metricgan} removes this constraint by combining the MetricGAN architecture with a non-intrusive neural assessment model, such as DNSMOS (for quality) or SRMR (for dereverberation). By training the discriminator to predict the score of a non-intrusive metric, the entire system can be optimized using only noisy speech. This represents a transformative shift, enabling the training of high-quality enhancement models on vast amounts of authentic, in-the-wild data, thereby improving real-world robustness and performance.

\subsection{Direct Optimization of Human Preference}

The ultimate goal is to optimize a system directly for subjective human preference. The human-in-the-loop paradigm achieves this by using a neural assessor trained on subjective data as the optimization target. The HL-StarGAN system~\cite{wang2024unsupervised} for enhancing face-masked speech illustrates this approach. Researchers first developed the MaskQSS assessor by collecting a database of face-masked speech and obtaining MOS ratings from human listeners. MaskQSS was trained to predict these MOS scores, and the enhancement model (generator) was then trained with a loss function that encouraged outputs predicted to achieve the highest possible MaskQSS score. After several iterations, we obtain the final HLStarGAN model. This two-stage training process creates a direct optimization pipeline that integrates the target subjective experience into the speech generation model, guided by the neural assessor.

\section{Neural Speech Assessment as a Decision Engine}
\subsection{Optimal Model Selection in an Ensemble System}

In ~\cite{zezario2019specialized} and ~\cite{zezario2021speech}, two novel speech enhancement systems were introduced that leverage neural speech assessment models as intelligent model selection frameworks to improve generalization, particularly in unseen conditions. Both approaches use Quality-Net to identify the optimal enhancement model for a given noisy utterance.

In ~\cite{zezario2019specialized}, the Specialized Speech Enhancement Model Selection (SSEMS) approach was proposed, in which specialized models are trained on data grouped by predefined attributes such as speaker gender and signal-to-noise ratio (SNR). During inference, all specialized models process the noisy input, and Quality-Net selects the output with the highest predicted quality.

In ~\cite{zezario2021speech}, the more advanced Zero-Shot Model Selection (ZMOS) framework was proposed. This data-driven approach applies zero-shot learning principles, using Quality-Net both to cluster training data via its latent “quality embeddings” and to perform model selection. This enables the system to choose the most suitable model for a given input without running all models, thereby improving efficiency.

Together, these works establish a powerful paradigm for adaptive speech enhancement, demonstrating that a learned, non-intrusive quality metric can serve as an effective runtime selection criterion, leading to significantly more robust performance across diverse noise conditions.

\subsection{Intelligibility-Aware  Beamforming System}
Beamforming frameworks typically rely on accurate estimation of the DOA, yet obtaining a reliable DOA is challenging, particularly under very low SIR conditions. In ~\cite{ting2023ians}, the Intelligibility-Aware Null-Steering (IANS) framework was proposed to address this challenge by optimally determining the DOA and enhancing speech intelligibility through beamforming.

IANS operates in two stages. First, a Null-Steering Beamformer (NSBF) generates multiple candidate signals by steering a suppression null across different angles. Second, a pre-trained deep learning model, STOI-Net, predicts the intelligibility of each candidate, and the system selects the signal with the highest predicted score—shifting the focus from conventional spatial filtering to direct intelligibility optimization.

Experimental results show that IANS significantly improves both intelligibility (STOI) and perceptual quality (PESQ) for noise-corrupted speech, achieving performance comparable to traditional beamformers with access to the true DOA of speech and noise. Moreover, IANS exhibits cross-lingual robustness, performing effectively on both English and Mandarin datasets without retraining STOI-Net. These results highlight direct intelligibility optimization as a powerful, language-independent alternative to conventional beamforming.

\section{Conclusion}
Neural speech assessment has transformed audio processing by bridging the gap between computational metrics and human perception. By serving as differentiable surrogates for complex objective measures and subjective judgments, models such as MetricGAN have enabled direct optimization for perceptual quality, achieving substantial gains over traditional approaches. These methods are now impacting a range of domains, including speech enhancement, text-to-speech, and voice conversion.

Despite this progress, key challenges remain. Generalization and calibration are major concerns, as models trained on subjective MOS data often underperform when applied to unseen conditions or systems. Multi-metric optimization is another frontier—future assessors must jointly account for multiple perceptual dimensions such as clarity, naturalness, and intelligibility, potentially through multiple discriminators or multi-objective training schemes. Interpretability and diagnostics are also pressing needs, allowing developers to understand why a model assigns certain quality scores and to obtain actionable feedback for improvement.

Looking ahead, the ultimate goal is personalization, where neural assessment models adapt to individual listener preferences and hearing profiles. Such systems could optimize output in real time for specific users, enabling hearing aids, communication platforms, and media services to deliver perceptually ideal audio. This shift from passive evaluation to active, user-specific optimization represents a pivotal step for next-generation audio technologies.

\bibliographystyle{IEEEtran}
\bibliography{mybib}

\begin{thebibliography}{10}
\providecommand{\url}[1]{#1}
\csname url@samestyle\endcsname
\providecommand{\newblock}{\relax}
\providecommand{\bibinfo}[2]{#2}
\providecommand{\BIBentrySTDinterwordspacing}{\spaceskip=0pt\relax}
\providecommand{\BIBentryALTinterwordstretchfactor}{4}
\providecommand{\BIBentryALTinterwordspacing}{\spaceskip=\fontdimen2\font plus
\BIBentryALTinterwordstretchfactor\fontdimen3\font minus \fontdimen4\font\relax}
\providecommand{\BIBforeignlanguage}[2]{{%
\expandafter\ifx\csname l@#1\endcsname\relax
\typeout{** WARNING: IEEEtran.bst: No hyphenation pattern has been}%
\typeout{** loaded for the language `#1'. Using the pattern for}%
\typeout{** the default language instead.}%
\else
\language=\csname l@#1\endcsname
\fi
#2}}
\providecommand{\BIBdecl}{\relax}
\BIBdecl

\bibitem{cooper2024review}
E.~Cooper, W.-C. Huang, Y.~Tsao, H.-M. Wang, T.~Toda, and J.~Yamagishi, ``A review on subjective and objective evaluation of synthetic speech,'' \emph{Acoustical Science and Technology}, vol.~45, no.~4, pp. 161--183, 2024.

\bibitem{reddy2021dnsmos}
C.~K. Reddy, V.~Gopal, and R.~Cutler, ``{DNSMOS}: A non-intrusive perceptual objective speech quality metric to evaluate noise suppressors,'' in \emph{Proc. ICASSP 2021}.

\bibitem{zezario2022deep}
R.~E. Zezario, S.-W. Fu, F.~Chen, C.-S. Fuh, H.-M. Wang, and Y.~Tsao, ``Deep learning-based non-intrusive multi-objective speech assessment model with cross-domain features,'' \emph{IEEE/ACM Transactions on Audio, Speech, and Language Processing}, vol.~31, pp. 54--70, 2022.

\bibitem{xu2023coded}
Z.~Xu, Z.~Zhao, and T.~Fingscheidt, ``Coded speech quality measurement by a non-intrusive {PESQ-DNN},'' \emph{IEEE/ACM Transactions on Audio, Speech, and Language Processing}, vol.~31, pp. 3404--3417, 2023.

\bibitem{saeki2024speechbertscore}
T.~Saeki, S.~Maiti, S.~Takamichi, S.~Watanabe, and H.~Saruwatari, ``{SpeechBERTScore}: Reference-aware automatic evaluation of speech generation leveraging {NLP} evaluation metrics,'' in \emph{Proc. INTERSPEECH 2024}.

\bibitem{fu2024self}
S.-W. Fu, K.-H. Hung, Y.~Tsao, and Y.-C.~F. Wang, ``Self-supervised speech quality estimation and enhancement using only clean speech,'' in \emph{Proc. ICLR 2024}.

\bibitem{fu2018quality}
S.-W. Fu, Y.~Tsao, H.-T. Hwang, and H.-M. Wang, ``{Quality-Net}: An end-to-end non-intrusive speech quality assessment model based on {BLSTM},'' in \emph{Proc. INTERSPEECH 2018}.

\bibitem{zezario2020stoi}
R.~E. Zezario, S.-W. Fu, C.-S. Fuh, Y.~Tsao, and H.-M. Wang, ``{STOI-Net}: A deep learning based non-intrusive speech intelligibility assessment model,'' in \emph{Proc. APSIPA ASC 2020}.

\bibitem{maiti2023speechlmscore}
S.~Maiti, Y.~Peng, T.~Saeki, and S.~Watanabe, ``{SpeechLMScore}: Evaluating speech generation using speech language model,'' in \emph{Proc. ICASSP 2023}.

\bibitem{lo2019mosnet}
C.-C. Lo, S.-W. Fu, W.-C. Huang, X.~Wang, J.~Yamagishi, Y.~Tsao, and H.-M. Wang, ``{MOSNet}: Deep learning based objective assessment for voice conversion,'' in \emph{Proc. INTERSPEECH 2019}.

\bibitem{leng2021mbnet}
Y.~Leng, X.~Tan, S.~Zhao, F.~Soong, X.-Y. Li, and T.~Qin, ``{MBNet}: {MOS} prediction for synthesized speech with mean-bias network,'' in \emph{Proc. ICASSP 2021}.

\bibitem{manocha2021noresqa}
P.~Manocha, B.~Xu, and A.~Kumar, ``{NORESQA}: A framework for speech quality assessment using non-matching references,'' in \emph{Proc. NeurIPS 2021}.

\bibitem{manocha2022speech}
P.~Manocha and A.~Kumar, ``Speech quality assessment through {MOS} using non-matching references,'' in \emph{Proc. INTERSPEECH 2022}.

\bibitem{huang2022ldnet}
W.-C. Huang, E.~Cooper, J.~Yamagishi, and T.~Toda, ``{LDNet}: Unified listener dependent modeling in {MOS} prediction for synthetic speech,'' in \emph{Proc. ICASSP 2022}.

\bibitem{cooper2022generalization}
E.~Cooper, W.-C. Huang, T.~Toda, and J.~Yamagishi, ``Generalization ability of {MOS} prediction networks,'' in \emph{Proc. ICASSP 2022}.

\bibitem{saeki2022utmos}
T.~Saeki, D.~Xin, W.~Nakata, T.~Koriyama, S.~Takamichi, and H.~Saruwatari, ``{UTMOS}: {Utokyo-SaruLab} system for {VoiceMOS Challenge} 2022,'' in \emph{Proc. INTERSPEECH 2022}.

\bibitem{maniati2022somos}
G.~Maniati, A.~Vioni, N.~Ellinas, K.~Nikitaras, K.~Klapsas, J.~S. Sung, G.~Jho, A.~Chalamandaris, and P.~Tsiakoulis, ``{SOMOS}: The samsung open {MOS} dataset for the evaluation of neural text-to-speech synthesis,'' in \emph{Proc. INTERSPEECH 2022}.

\bibitem{qi2023ssl}
Z.~Qi, X.~Hu, W.~Zhou, S.~Li, H.~Wu, J.~Lu, and X.~Xu, ``{LE-SSL-MOS}: Self-supervised learning {MOS} prediction with listener enhancement,'' in \emph{Proc. ASRU 2023}.

\bibitem{minixhofer2025ttsds2}
C.~Minixhofer, O.~Klejch, and P.~Bell, ``{TTSDS2}: Resources and benchmark for evaluating human-quality text to speech systems,'' \emph{arXiv preprint arXiv:2506.19441}, 2025.

\bibitem{huang2022voicemos}
W.-C. Huang, E.~Cooper, Y.~Tsao, H.-M. Wang, T.~Toda, and J.~Yamagishi, ``The {VoiceMOS Challenge} 2022,'' in \emph{Proc. INTERSPEECH 2022}.

\bibitem{cooper2023voicemos}
E.~Cooper, W.-C. Huang, Y.~Tsao, H.-M. Wang, T.~Toda, and J.~Yamagishi, ``The {VoiceMOS Challenge} 2023: Zero-shot subjective speech quality prediction for multiple domains,'' in \emph{Proc. ASRU 2023}.

\bibitem{huang2024voicemos}
W.-C. Huang, S.-W. Fu, E.~Cooper, R.~E. Zezario, T.~Toda, H.-M. Wang, J.~Yamagishi, and Y.~Tsao, ``The {VoiceMOS Challenge} 2024: Beyond speech quality prediction,'' in \emph{Proc. SLT 2024}.

\bibitem{zhang2024urgent}
W.~Zhang, R.~Scheibler, K.~Saijo, S.~Cornell, C.~Li, Z.~Ni, A.~Kumar, J.~Pirklbauer, M.~Sach, S.~Watanabe \emph{et~al.}, ``{URGENT Challenge}: Universality, robustness, and generalizability for speech enhancement,'' in \emph{Proc. INTERSPEECH 2024}.

\bibitem{blanco2023avse}
A.~L.~A. Blanco, C.~Valentini-Botinhao, O.~Klejch, M.~Gogate, K.~Dashtipour, A.~Hussain, and P.~Bell, ``{AVSE Challenge}: Audio-visual speech enhancement challenge,'' in \emph{Proc. SLT 2023}.

\bibitem{wang2025qualispeech}
S.~Wang, W.~Yu, X.~Chen, X.~Tian, J.~Zhang, L.~Lu, Y.~Tsao, J.~Yamagishi, Y.~Wang, and C.~Zhang, ``Qualispeech: A speech quality assessment dataset with natural language reasoning and descriptions,'' in \emph{Proc. ACL 2025}.

\bibitem{ahmed2025study}
S.~Ahmed, R.~E. Zezario, N.~Saleem, A.~Hussain, H.-M. Wang, and Y.~Tsao, ``A study on speech assessment with visual cues,'' in \emph{Proc. INTERSPEECH 2025}.

\bibitem{parmonangan2023common}
I.~H. Parmonangan, ``Common brain activity features discretization for predicting perceived speech quality,'' \emph{Procedia Computer Science}, vol. 216, pp. 774--783, 2023.

\bibitem{hsin2025exploring}
C.-H. Hsin, C.-Y. Lee, and Y.~Tsao, ``Exploring {N400} predictability effects during sustained speech comprehension: From listening-related fatigue to speech enhancement evaluation,'' \emph{Ear and Hearing}, vol.~46, no.~4, pp. 922--940, 2025.

\bibitem{fu2019metricgan}
S.-W. Fu, C.-F. Liao, Y.~Tsao, and S.-D. Lin, ``{MetricGAN}: Generative adversarial networks based black-box metric scores optimization for speech enhancement,'' in \emph{Proc. ICML 2019}.

\bibitem{nayem2021incorporating}
K.~M. Nayem and D.~S. Williamson, ``Incorporating embedding vectors from a human mean-opinion score prediction model for monaural speech enhancement.'' in \emph{Proc. INTERSPEECH 2021}.

\bibitem{xu2021deep}
Z.~Xu, M.~Strake, and T.~Fingscheidt, ``Deep noise suppression with non-intrusive {PESQNet} supervision enabling the use of real training data,'' in \emph{Proc. INTERSPEECH 2021}.

\bibitem{fu2022metricgan}
S.-W. Fu, C.~Yu, K.-H. Hung, M.~Ravanelli, and Y.~Tsao, ``{MetricGAN-U}: Unsupervised speech enhancement/dereverberation based only on noisy/reverberated speech,'' in \emph{Proc. ICASSP 2022}.

\bibitem{xu2022deep}
Z.~Xu, M.~Strake, and T.~Fingscheidt, ``Deep noise suppression maximizing non-differentiable pesq mediated by a non-intrusive {PESQNet},'' \emph{IEEE/ACM Transactions on Audio, Speech, and Language Processing}, vol.~30, pp. 1572--1585, 2022.

\bibitem{ting2023ians}
W.-Y. Ting, S.-S. Wang, Y.~Tsao, and B.~Su, ``{IANS}: Intelligibility-aware null-steering beamforming for dual-microphone arrays,'' in \emph{Proc. MLSP 2023}.

\bibitem{nayem2023attention}
K.~M. Nayem and D.~S. Williamson, ``Attention-based speech enhancement using human quality perception modeling,'' \emph{IEEE/ACM Transactions on Audio, Speech, and Language Processing}, vol.~32, pp. 250--260, 2023.

\bibitem{wang2024unsupervised}
S.-S. Wang, J.-Y. Chen, B.-R. Bai, S.-H. Fang, and Y.~Tsao, ``Unsupervised face-mask speech enhancement using generative adversarial networks with human-in-the-loop assessment metrics,'' \emph{IEEE/ACM Transactions on Audio, Speech, and Language Processing}, vol.~32, pp. 3826--3837, 2024.

\bibitem{chao2024investigation}
R.~Chao, W.-H. Cheng, M.~La~Quatra, S.~M. Siniscalchi, C.-H.~H. Yang, S.-W. Fu, and Y.~Tsao, ``An investigation of incorporating {M}amba for speech enhancement,'' in \emph{Proc. SLT 2024}.

\bibitem{wu2025reexamining}
H.~Wu, A.~Aroudi, B.~Xu, A.~Pandey, F.~Nesta, A.~Kumar, A.~Reich, and K.~Tan, ``Reexamining the efficacy of {MetricGAN} for speech enhancement,'' in \emph{Proc. ICASSP 2025}.

\bibitem{lu2025explicit}
Y.-X. Lu, Y.~Ai, and Z.-H. Ling, ``Explicit estimation of magnitude and phase spectra in parallel for high-quality speech enhancement,'' \emph{Neural Networks}, p. 107562, 2025.

\bibitem{jiang2025mos}
W.~Jiang, F.~Wen, and K.~Yu, ``{MOS-GAN}: Mean opinion score {GAN} for unsupervised speech enhancement,'' \emph{IEEE Signal Processing Letters}, 2025.

\bibitem{zezario2019specialized}
R.~E. Zezario, S.-W. Fu, X.~Lu, H.-M. Wang, Y.~Tsao \emph{et~al.}, ``Specialized speech enhancement model selection based on learned non-intrusive quality assessment metric.'' in \emph{Proc. INTERSPEECH 2019}.

\bibitem{zezario2021speech}
R.~E. Zezario, C.-S. Fuh, H.-M. Wang, and Y.~Tsao, ``Speech enhancement with zero-shot model selection,'' in \emph{Proc. EUSIPCO 2021}.

\bibitem{rix2001perceptual}
A.~W. Rix, J.~G. Beerends, M.~P. Hollier, and A.~P. Hekstra, ``Perceptual evaluation of speech quality ({PESQ})-a new method for speech quality assessment of telephone networks and codecs,'' in \emph{Proc. ICASSP 2001}.

\bibitem{gaoxiong2012perceptual}
Y.~Gaoxiong and Z.~Wei, ``The perceptual objective listening quality assessment algorithm in telecommunication: Introduction of {ITU-T} new metrics {POLQA},'' in \emph{Proc. ICCC 2012}.

\bibitem{fu2021metricgan+}
S.-W. Fu, C.~Yu, T.-A. Hsieh, P.~Plantinga, M.~Ravanelli, X.~Lu, and Y.~Tsao, ``{MetricGAN+}: An improved version of {MetricGAN} for speech enhancement,'' in \emph{Proc. INTERSPEECH 2021}.

\bibitem{cao2022cmgan}
R.~Cao, S.~Abdulatif, and B.~Yang, ``{CMGAN}: {Conformer}-based {Metric-GAN} for speech enhancement,'' \emph{Proc. INTERSPEECH 2022}.

\bibitem{close2022metricgan+}
G.~Close, T.~Hain, and S.~Goetze, ``{MetricGAN}+/-: Increasing robustness of noise reduction on unseen data,'' in \emph{Proc. EUSIPCO 2022}.

\bibitem{cheng2023speech}
J.~Cheng, R.~Liang, L.~Zhao, C.~Huang, and B.~W. Schuller, ``Speech denoising and compensation for hearing aids using an {FTCRN-based metric GAN},'' \emph{IEEE Signal Processing Letters}, vol.~30, pp. 374--378, 2023.

\bibitem{shin2023metricgan}
W.~Shin, B.~H. Lee, J.~S. Kim, H.~J. Park, and S.~W. Han, ``{MetricGAN-OKD}: multi-metric optimization of {MetricGAN} via online knowledge distillation for speech enhancement,'' in \emph{Proc. ICML 2023}.

\bibitem{hou2023convolutional}
Z.~Hou, Q.~Hu, T.~Sun, Y.~Hu, C.~Zhu, and K.~Chen, ``Convolutional recurrent {MetricGAN} with spectral dimension compression for full-band speech enhancement,'' in \emph{Proc. ICASSP 2023}.

\bibitem{mai2025metricgan+}
Y.~Mai and S.~Goetze, ``{MetricGAN+KAN}: {Kolmogorov-Arnold} networks in metric-driven speech enhancement systems,'' in \emph{Proc. ICASSP 2025}.

\bibitem{taal2011algorithm}
C.~H. Taal, R.~C. Hendriks, R.~Heusdens, and J.~Jensen, ``An algorithm for intelligibility prediction of time--frequency weighted noisy speech,'' \emph{IEEE Transactions on Audio, Speech, and Language Processing}, vol.~19, no.~7, pp. 2125--2136, 2011.

\bibitem{goodfellow2020generative}
I.~Goodfellow, J.~Pouget-Abadie, M.~Mirza, B.~Xu, D.~Warde-Farley, S.~Ozair, A.~Courville, and Y.~Bengio, ``Generative adversarial networks,'' \emph{Communications of the ACM}, vol.~63, no.~11, pp. 139--144, 2020.

\end{thebibliography}

\end{document}